\documentclass[twocolumn,showpacs,preprintnumbers,amsmath,amssymb,pre]{revtex4-1}
\usepackage{graphicx}
\usepackage{dcolumn}
\usepackage{bm}
\usepackage{epstopdf}
\usepackage{ulem}

\usepackage{color}

\begin{document}

\title{
Projective-truncation-approximation study of the one-dimensional $\phi^{4}$ lattice model}

\author{Kou-Han Ma}
\affiliation{Department of Physics, Renmin University of China, 100872 Beijing, China}

\author{Yan-Jiang Guo}
\affiliation{Department of Physics, Renmin University of China, 100872 Beijing, China}

\author{Lei Wang}
\affiliation{Department of Physics, Renmin University of China, 100872 Beijing, China}

\author{Ning-Hua Tong}
\email{nhtong@ruc.edu.cn}
\affiliation{Department of Physics, Renmin University of China, 100872 Beijing, China}
\date{\today}

\begin{abstract}
In this paper, we first develop the projective truncation approximation (PTA) in the Green's function equation of motion (EOM) formalism for classical statistical models. To implement PTA for a given Hamiltonian, we choose a set of basis variables and projectively truncate the hierarchical EOM. We apply PTA to the one-dimensional $\phi^{4}$ lattice model. Phonon dispersion and static correlation functions are studied in detail. Using one- and two-dimensional bases, we obtain results identical to and beyond the quadratic variational approximation, respectively. In particular, we analyze the power-law temperature dependence of the static averages in the low- and high- temperature limits, and we give exact exponents.

\end{abstract}


\maketitle

\begin{section}{Introduction}

Classical many-body systems are an important research topic in condensed matter physics, covering such diverse subjects as state equation of atomic/molecular gases\cite{JOH1}, glass formation in liquid\cite{PGD1}, anomalous heat conductivity in low dimensional atomic chains\cite{SL1}, and structural phase transition\cite{MF1}, {\it etc}. Accurate and efficient solution of the related classical statistical models play a central role in the theoretical study. Modern computer-based techniques such as Monte Carlo and molecular dynamics are powerful but not sufficient to solve all the problems, due to the limitations from the computational complexity in size and time. Analytical methods, such as mode coupling theory\cite{SPD1}, renormalization group\cite{TY2}, variational method\cite{TD3,BL4}, {\it etc}, are still extensively used in the study. The present work is an effort to promote one of the analytical methods, Green's function (GF) equation of motion (EOM), to an advanced level. We apply it to the study of one-dimensional $\phi^{4}$ lattice model for interacting particles on a chain. By enlarging the size of the variable basis, we obtain improved phonon dispersion and static averages, demonstrating the applicability of the proposed method to classical statistical models with continuous variables. Qualitatively accurate temperature dependence behavior in the low and high temperature limit can be extracted from our analysis.

The formalism of EOM of double time GF has a long history. It was developed first for quantum system in 1950s\cite{Martin1,Bogoliubov1,Tyablikov1,Zubarev1} and then generalized to classical systems by Bogoliubov and Sadovnikov using a variational technique\cite{Bogoliubov2}. Herzel rederived the EOM of classical GF\cite{Herzel1} using the double time theory of Rostoker\cite{Rostoker1} and the Heisenberg picture for classical statistics\cite{Aronson1}. A many-time GF and resolvent formalism of classical GF was subsequently developed by Herzel\cite{Herzel3}. The meaning of these GFs as linear and higher order response coefficients to external time-dependent perturbation was elaborated in Refs.\cite{Herzel2} and \cite{Tanaka1}. Applying this method to ideal gas, Smith obtained the exact density-density correlation function\cite{Smith1}. Campana {\it et al.} introduces the spectral function of the classical GF and proved the spectral theorem\cite{Campana1}. A closely related method, the spectral density method\cite{Kalashnikov1}, was transplanted from quantum systems to classical systems and was applied to a variety of classical many-body statistical problems\cite{Campana1,Cavallo1,Cavallo2,Cavallo3}. A Callen-type decoupling truncation of the hierarchical EOM was carried out for the classical Heisenberg model\cite{Campana2,Campana3}.

In EOM method, a lower-order GF is related to a higher-order one and so on, until at some point this chain has to be truncated to form closed equations for GFs\cite{YLL1}. Traditional truncation procedures often rely heavily on physical intuition and are difficult to generalize. Certain analytical requirements of GFs, such as sum rule, positiveness of spectral weight, and real simple poles, are hard to guarantee by truncation approximations. Besides, the chain of EOM will involve many averages, which are usually calculated self-consistently from the GFs via the fluctuation-dissipation theorem. Due to truncation, the number of unknowns could exceed the number of equations and some additional approximations need to be invoked. All these make the traditional truncation approximation of EOM a poorly controlled method.

Based on the idea of operator projection\cite{Mori,Zwanzig}, a projective truncation approximation (PTA) was developed for quantum systems\cite{Fan1} to overcome the shortcomings of the traditional truncation approximation mentioned above. In this work, we adopt the same idea and develop PTA for the classical statistical models.
We apply PTA to the study of one-dimensional $\phi^4$ lattice model\cite{KA32,BH33}, both to demonstrate the applicability of the method and to disclose the underlying physics of this model. This model has been the focus of a series studies in the context of low-dimensional heat transport\cite{KA32, BH33, KA31.1, KA34, AD35, NL36, LX37, NL38} and chaotic dynamics\cite{WGH36.1, KA36.2}.
The existence of quartic potential in this model opens a gap in the phonon spectrum at finite temperature and leads to normal heat transport behavior. The phonon dispersion has been analyzed by various methods, such as the theories of self-consistent phonon (i.e. quadratic variational method)\cite{JL40}, effective phonon\cite{EP1}, anharmonic phonon\cite{AP1}, and the resonance phonon\cite{LX37}. Among them, the first one is an analytical method and the latter three require numerical results as input.

In this study, we focus on the phonon dispersion and static averages of this model. To implement PTA, we need to choose a set of basis variables to projectively truncate the EOM. Using one- and two-dimensional bases within PTA, respectively, we obtain results identical to and beyond those from the variational method with quadratic reference Hamiltonian, respectively. Our method provides a new way to calculate the phonon spectrum of nonlinear lattice systems. The temperature-dependence of static averages are also analyzed. We argue that the obtained asymptotic low and high temperature power law are qualitatively exact.

The rest of this paper is arranged as follows. For completeness, we first review the formalism of GF EOM for classical systems in Sec. II. In Sec. III, we develop the formalism of PTA for classical system.
In Sec. IV, we apply PTA to one-dimensional $\phi^4$ lattice model and summarize the formulas. Section V is denoted to the discussion of PTA results under different bases. A summary and discussion are given in Sec. VI.

\end{section}

\begin{section}{Double Time Green's Function Equation of Motion}

In this section, we give a pedagogic review of the GF EOM for classical system, setting up the frame for PTA in the next section. A complete discussion can be found in Ref.\cite{Cavallo2}. Compared to existing formalism of EOM\cite{Herzel1,Cavallo2}, in this work, the fluctuation-dissipation theorem is modified such that it is applicable to variables with finite static component.

Suppose we have a classical system with canonical ordinates $(q_1, q_2, ..., q_N)$ and momenta $(p_1, p_2, ..., p_N)$. The Hamiltonian $H(q, p)$ describes a conserving system without dissipative forces. Here and below, we will use $q$ and $p$ as the short-handed notion for $(q_1, q_2, ..., q_N)$ and $(p_1, p_2, ..., p_N)$, respectively. In this paper, we only consider Hamiltonian and dynamical variables that do not explicitly contain time $t$, such as $A(q, p)$ and $B(q, p)$, {\it etc}.
The time evolution of $q(t)$ and $p(t)$ is determined by Hamilton's equations
\begin{eqnarray}      \label{eq.2}
&&   \frac{d q_i(t)}{dt} = \frac{\partial H(q,p)}{\partial p_i} \big|_{q=q(t), \,\,p=p(t)} , \nonumber \\
  &&  \frac{d p_i(t)}{dt} = - \frac{\partial H(q,p) }{\partial q_i}\big|_{q=q(t), \,\,p=p(t)}.
\end{eqnarray}
The Poisson bracket between two variables $A$ and $B$ is defined as
\begin{eqnarray}      \label{eq.1}
&&  \{ A(q, p), B(q, p) \}  \nonumber \\
&&  \equiv \sum_{i=1}^{N} \left[ \frac{\partial A(q, p)}{\partial q_i} \frac{\partial B(q, p)}{\partial p_i} - \frac{\partial A(q, p)}{\partial p_i} \frac{\partial B(q, p)}{\partial q_i} \right] \nonumber \\
&& = \frac{\partial A(q, p)}{\partial q} \frac{\partial B(q, p)}{\partial p} - \frac{\partial A(q, p)}{\partial p} \frac{\partial B(q, p)}{\partial q}.
\end{eqnarray}
In the following, as in the third line of Eq.(\ref{eq.1}), we will neglect the summation over $i$ and  abbreviate $q_i$ and $p_i$ by $q$ and $p$.
The standard Poisson brackets $\{ q_i, p_j \} = \delta_{ij}$ and $\{ q_i, q_j \} = \{ p_i, p_j \} = 0$ are special cases of Eq.(\ref{eq.1}). It is noted that the Poisson bracket defined above is invariant under canonical transformation\cite{Goldstein1}. That is,
\begin{eqnarray}      \label{eq.1.1}
&&  \{ A(q, p), B(q, p) \}  \nonumber \\
&& = \frac{\partial A(q, p)}{\partial Q} \frac{\partial B(q, p)}{\partial P} - \frac{\partial A(q, p)}{\partial P} \frac{\partial B(q, p)}{\partial Q} ,  \nonumber \\
 &&
\end{eqnarray}
with $Q_i = Q_i(q,p)$, $P_i = P_i(q,p)$ being canonical transformation.

In terms of the Poisson bracket, the time evolution of $A(t)=A\left[q(t), p(t)\right]$ obeys the EOM
\begin{eqnarray}      \label{eq.3}
 \frac{d}{dt} A[q(t),p(t)] = \{ A(q,p), H(q,p) \}(t).
\end{eqnarray}
Under this equation the energy is conserved $dH[q(t), p(t)]/dt = 0$.
Since $A[q(t), p(t)]$ follows a deterministic equation Eq.(\ref{eq.3}), we have an alternative representation for it, $ A[q(t), p(t)] = A[q(0), p(0); t]$. This change of representation is actually a transition from the Schr\"{o}dinger picture, where the state $\Gamma(t)=(q(t),p(t))$ evolves with time and the operator $A(q,p)$ does not, to the Heisenberg picture where the state stays at $\Gamma(0)=(q(0),p(0))$ while the operators evolve\cite{Aronson1,Herzel2}. At $t=0$, the two pictures coincide.

The retarded Green's function of two dynamical variables $A[q(t), p(t)]$ and $B[q(t^{\prime}), p(t^{\prime})]$ is defined as\cite{Herzel1,Smith1,Cavallo1,Cavallo2}
\begin{equation}       \label{eq.11.1}
   G^{r}\left[ A(t)|B(t^{\prime}) \right] \equiv \theta(t-t^{\prime}) \langle \{ A(t), B(t^{\prime}) \} \rangle.
\end{equation}
Here $\theta(x)$ is the Heaviside step function. $\langle O \rangle$ is the average of variable $O$ in equilibrium state. $\{...\}$ is the Poission bracket defined in Eq.(\ref{eq.1}). Eq.(\ref{eq.11.1}) gives the linear response coefficient of $\langle A(t) \rangle$ under a weak perturbation proportional to $B(t^{\prime})$\cite{Tanaka1,Herzel2}.
Some remarks about the definition Eq.(\ref{eq.11.1}) are in order.
For variables at unequal times $A((t)$ and $B(t^{\prime})$, it is more convenient to use the Poisson bracket Eq.({\ref{eq.1.1}}) and choose a special set of canonical variables, $Q_i(q,p)=q_i(0)$, $P_i(q,p)=p_i(0)$. $\{ A(t), B(t^{\prime}) \}$ in Eq.(\ref{eq.11.1}) is then written in the Heisenberg picture as
\begin{eqnarray}      \label{eq.5.5}
 &&  \{ A(t), B(t^{\prime}) \}  \nonumber \\
 & = & \frac{\partial A[q(0),p(0); t]}{ \partial q(0)}\frac{\partial B[q(0),p(0); t^{\prime}]}{ \partial p(0)} \nonumber \\
 && - \frac{\partial A[q(0),p(0); t]}{ \partial p(0)} \frac{\partial B[q(0),p(0); t^{\prime}]}{ \partial q(0)}.
\end{eqnarray}
Besides the usual properties of Poinsson bracket such as Jacobi's identity,
\begin{eqnarray}     \label{eq.5.13.5}
  && \{ A(t_1), \{ B(t_2), C(t_3) \} \} +  \{B(t_2), \{ C(t_3), A(t_1) \} \} \nonumber \\
  &+& \{ C(t_3), \{ A(t_1), B(t_2) \} \} =  0,
\end{eqnarray}
Eq.(\ref{eq.5.5}) also has the following notable properties,
\begin{equation}     \label{eq.5.12}
   \frac{\partial}{\partial t} \{ A(t), B(t^{\prime}) \} =   \left\{ \frac{\partial}{\partial t} A(t), B(t^{\prime}) \right \};
\end{equation}
and the cyclic relation
\begin{eqnarray}     \label{eq.5.13}
  && \int dq \int dp A(t_1) \{ B(t_2), C(t_3) \}  \nonumber \\
 &&=   \int dq \int dp B(t_2) \{ C(t_3), A(t_1) \}.
\end{eqnarray}
Eq.(\ref{eq.5.13}) can be obtained from integrating by part and neglecting the boundary term. It is valid when one of the operators among $A(t_1)$, $B(t_2)$, and $C(t_3)$ becomes zero at the boundary of the phase space. In particular, it holds when $A(t_1) = e^{-\beta H(q,p)}/Z$ is the equilibrium density operator.

In Gibbs statistical theory, a state of the studied system is described by the probability density $\rho(q,p,t)$ of the ensemble distribution. The time evolution of $\rho(q,p,t)$ is governed by the Liouville equation
\begin{equation}   \label{eq.5.14}
   \frac{d}{dt} \rho[q(t), p(t), t] = 0.
\end{equation}
In Schr\"{o}dinger picture, the ensemble average of a physical quantity $O(q,p)$ is given as
\begin{equation}   \label{eq.5.16}
  \langle O \rangle (t) \equiv  \int dq \int dp O(q,p) \rho(q,p,t).
\end{equation}
Using the invariance of phase space volume $dq(0)dp(0) = dq(t) dp(t)$ and Liouville theorem $d\rho[q(t), p(t), t]/dt=0$, we obtain the expression in the Heisenberg picture,
\begin{eqnarray}  \label{eq.5.18}
   \langle O \rangle (t)  = \int dq(0) \int dp(0) O[q(0), p(0); t] \rho[q(0), p(0), 0]. \nonumber \\
    &&
\end{eqnarray}
Eq.(\ref{eq.5.18}) says that the ensemble average of $O(t)$ can be calculated by averaging $O[q(0), p(0), t]$ over the initial distribution of $q(0)$ and $p(0)$. This formalism is used in the definition of Green's function in Eq.(\ref{eq.11.1}), with the equilibrium state $\rho(q,p,t)$ (considering canonical ensemble here)
\begin{equation}     \label{eq.6}
  \rho(q, p, t) = \frac{1}{Z} e^{-\beta H(q,p)}.
\end{equation}
The partition function is $Z = \int dq \int dp \exp{[-\beta H(q,p)] }$. Here and below, we neglect the factor $1/(N!h^{N})$ for brevity. $\beta = 1/(kT)$ is the inverse temperature. It is easy to prove the time-translation invariance of equilibrium state averages, $\langle O(t)  \rangle =  \langle O \rangle$,  $\langle A(t)B(t^{\prime}) \rangle = \langle A(t-\tau)B(t^{\prime}-\tau) \rangle$, and $\langle \{A(t), B(t^{\prime}) \} \rangle = \langle \{ A(t-\tau), B(t^{\prime}-\tau) \} \rangle$.
Taking derivative of $t$ on both sides of $\langle O(t)  \rangle =  \langle O \rangle$, we obtain an important conservation relation for arbitrary operator $O(q,p)$,
\begin{equation}    \label{eq.8.1}
\langle \{ O(q,p), H(q,p) \} \rangle = 0.
\end{equation}
This equation has the virial identity $\langle \triangledown \cdot \vec{f}(q) \rangle = \beta \langle \vec{f}(q) \cdot \triangledown H \rangle$ as a special case\cite{BJ1}. Here, $\vec{f}(q)$ is a polynomial function of $q$. Letting $O(q,p)=q_ip_i$, we also obtain the generalized equipartition theorem $\langle q_i \partial H/\partial q_i \rangle = T$. It will be used to simplify the EOM and to analyze the properties of physical quantities in the low- and high- temperature limits for the one-dimensional $\phi^4$ lattice model. The cyclic relation Eq.(\ref{eq.5.13}) implies
\begin{eqnarray}    \label{eq.8.3}
 \langle \{ A(t), B(t^{\prime})  \} \rangle &=& \beta \langle \{ A, H \}(t) B(t^{\prime}) \rangle  \nonumber \\
&=& -\beta \langle A(t) \{ B, H \}(t^{\prime}) \rangle  .
\end{eqnarray}

Let us now derive the EOM for  $G^{r}\left[ A(t)|B(t^{\prime}) \right]$. We do derivative with respect to $t$ on both sides of Eq.(\ref{eq.11.1}) and employ Eqs.(\ref{eq.3}) and (\ref{eq.5.12}). Note that in the Heisenberg picture where $A[q(t), p(t)] = A[q(0), p(0); t]$, we have $ \partial{A(t)} / \partial{t} =dA(t) / dt$. We obtain
\begin{eqnarray}       \label{eq.14}
  && \frac{\partial}{\partial t} G^{r}\left[ A(t)|B(t^{\prime}) \right] \nonumber \\
  && = \delta(t-t^{\prime}) \langle \{ A, B\} \rangle + G^{r}\left[ \{A, H\}(t)|B(t^{\prime}) \right].
\end{eqnarray}
The Fourier transformation of GF is defined as
\begin{equation}      \label{eq.15.1}
   G^{r}(A|B)_{\omega} = \int_{-\infty}^{\infty}  G^{r}\left[ A(t)|B(t^{\prime}) \right] e^{i (t-t^{\prime} ) (\omega + i \eta)} d(t-t^{\prime}).
\end{equation}
$\eta$ is an infinitesimal positive number. $G^{r}(A|B)_{\omega}$ is a function of $\omega + i \eta$. Combining Eqs.\eqref{eq.14} and \eqref{eq.15.1}, we obtain the EOM for retarded GF in frequency domain,
\begin{equation}      \label{eq.15.2}
   (\omega+i\eta) G^{r}( A | B )_{\omega} = i \langle \{ A, B \} \rangle + i G^{r}( \{ A, H \} | B )_{\omega}.
\end{equation}
This equation is usually expressed in a more compact form, i.e., the EOM of Zubarev GF $G(A|B)_{\omega}$ (without the superscript r, obtained by substituting the argument $\omega + i\eta$ of $G^{r}(A|B)_{\omega}$ by $\omega$)\cite{Zubarev1}. It reads
\begin{equation}      \label{eq.16}
   \omega G( A | B )_{\omega} = i \langle \{ A, B \} \rangle + i G( \{ A, H \} | B )_{\omega}.
\end{equation}
The retarded GF $G^{r}(A|B)_{\omega}$ can be recovered by analytical continuation of the Zubarev GF $G(A|B)_{\omega}$, i.e., $G^{r}(A|B)_{\omega} = G( A | B)_{\omega\to\omega + i \eta}$.
Similarly, derivative of Eq.(\ref{eq.11.1}) with respect to $t^{\prime}$ gives the right-hand side EOM
\begin{equation}      \label{eq.17}
   \omega G(A|B)_{\omega} = i \langle \{ A, B \} \rangle - i G( A | \{B, H \} )_{\omega}.
\end{equation}

The static averages of equilibrium state can be obtained from the corresponding GF via the fluctuation-dissipation theorem\cite{Herzel1,Cavallo2}
\begin{equation}      \label{eq.18.1}
   \langle A B \rangle = \frac{1}{\beta} \int_{-\infty}^{\infty} \frac{\Lambda_{A,B}(\omega)}{\omega} d\omega + \langle A_0 B_0 \rangle.
\end{equation}
Here, $A_0$ and $B_0$ are the zero-frequency components of $A(t)$ and $B(t)$, respectively. The precise definition and some properties of the zero-frequency component
 $X_0$ of a general variable $X$ is summarized in Appendix B. The spectral function $\Lambda_{A,B}(\omega)$ in the above equation is defined as\cite{Herzel1,Cavallo2}
\begin{equation}      \label{eq.19.1}
   \Lambda_{A,B}(\omega)  \equiv  \frac{i}{2\pi} \left[ G( A|B )_{\omega+ i\eta} - G( A|B)_{\omega-i\eta} \right].
\end{equation}
The proof of Eq.(\ref{eq.18.1}) is given in Appendices A and B.

Note that besides a factor $2\pi$ difference in the definition, Eq.(\ref{eq.18.1}) is different from previous works\cite{Herzel1,Cavallo2} in that the contribution from static components of $A(t)$ and $B(t)$ are singled out. This equation has a wider application range than those in Refs.\cite{Herzel1} and \cite{Cavallo2}. In the case that $A_0$ or $B_0$ is a constant number, $\langle A_0 B_0\rangle = \langle A \rangle \langle B \rangle$. In general, $A_0$ and $B_0$ are conserving quantities with possibly nonzero statistical fluctuations and computing $\langle A_0 B_0 \rangle$ is a nontrivial task. This problem also arises in the commutator GF EOM formalism for quantum systems. Possible solutions are discussed in literatures \cite{Stevens1,Ramos1,Froebrich1,Froebrich2}. Similar methods can be used here to compute $\langle A_0 B_0\rangle$ in the classical GF EOM formalism. For the $\Phi^4$ lattice model that we will study in this paper, $\langle A_0 B_0 \rangle = 0$ (see below).

\end{section}

\begin{section}{Projective Truncation Approximation}

The above formalism of GF EOM is standard and has been obtained in previous literatures.
In this section, we present the new development of this work, i.e., introducing PTA into the GF EOM forr classical systems. PTA was proposed by Fan {\it et al.} first for quantum GF EOM\cite{Fan1}. It is a systematic method for truncating the EOM and it has controllable precision\cite{Fan2}. Recently, this method is used in the study of phase diagram of two dimensional spinless fermion model\cite{Ma1}. Given the similar structure of EOM in quantum and classical cases, PTA can well be transplanted to classical GF EOM, with special structure of classical system taken into account.

We first generalize the GF EOM formalism to a matrix form. Suppose we have a vector of basis variables $\vec{A} = (A_1, A_2, ..., A_n)^{T}$ which are in general complex. We assume a real Hamiltonian $H$ and that the coordinates $q_i$ and momenta ${p_i}$ can be canonically transformed into real variables. Due to the invariance of Poisson bracket under canonical transformation of variables, the formula in previous section still applies to complex variables $\{ A_i\}$. We have $\langle O \rangle^{\ast} = \langle O^{\ast} \rangle$ and $\{X(t), Y(t^{\prime}) \}^{\ast} = \{ X^{\ast}(t), Y^{\ast}(t^{\prime}) \}$.

The matrix of retarded GF is defined as
\begin{equation}      \label{eq.20.1}
   G^{r}\left( \vec{A}(t) \big| \vec{A}^{\dag}(t^{\prime}) \right) \equiv \theta(t-t^{\prime}) \left\langle  \left\{\vec{A}(t), \vec{A}^{\dag}(t^{\prime}) \right\} \right\rangle.
\end{equation}
The Fourier transformation of GF and the spectral density function are given respectively as
\begin{eqnarray}      \label{eq.20.2}
&&  G^{r} ( \vec{A} \big| \vec{A}^{\dag} )_{\omega} = \int_{-\infty}^{\infty} d(t-t^{\prime}) G^{r} \left[ \vec{A}(t) \big| \vec{A}^{\dag}(t^{\prime}) \right] e^{i(\omega + i\eta)(t-t^{\prime})},   \nonumber \\
&&
\end{eqnarray}
and
\begin{eqnarray}      \label{eq.20.3}
 \Lambda_{ \vec{A}, \vec{A}^{\dag} }(\omega) = \frac{i}{2 \pi} \left[ G( \vec{A}\big| \vec{A}^{\dag} )_{\omega + i \eta} - G( \vec{A}\big| \vec{A}^{\dag})_{\omega - i \eta} \right].
\end{eqnarray}
The fluctuation-dissipation theorem is generalized into
\begin{equation}      \label{eq.20.4}
 {\bf C} =  \frac{1}{\beta} \int_{-\infty}^{\infty} d\omega \frac{\Lambda_{ \vec{A}, \vec{A}^{\dag} }(\omega)}{\omega} + {\bf C}_0.
\end{equation}
Here, the correlation matrix ${\bf C}_{n \times n}$ has the element ${\bf C}_{ij} = \langle A^{\ast}_i A_j \rangle$. $({\bf C}_0)_{ij} = \langle A^{\ast}_{i0} A_{j0} \rangle$ is the correlation of zero-frequency components $A_{i0}$ and $A_{j0}$ of the basis variables $A_i$ and $A_j$. ${\bf C}$ and ${\bf C}_0$ are both Hermitian and positive definite matrices. 
The two EOMs of the GF matrix read
\begin{equation}      \label{eq.20.5}
   \omega G( \vec{A} \big| \vec{A}^{\dag} )_{\omega} = i \langle \{\vec{A}, \vec{A}^{\dag} \}\rangle + i G( \{ \vec{A}, H\} \big| \vec{A}^{\dag} )_{\omega},
\end{equation}
and
\begin{equation}      \label{eq.20.6}
   \omega G( \vec{A}\big| \vec{A}^{\dag} )_{\omega} = i \langle \{\vec{A}, \vec{A}^{\dag} \}\rangle - i G( \vec{A} \big| \{ \vec{A}^{\dag}, H \} )_{\omega}.
\end{equation}

Before making PTA, we first invoke a special feature of the classical dynamics. It has been observed that the classical GF always has poles in plus-and-minus pairs\cite{Cavallo2,Prigogine1}. This reminds us that there could be some structure in the Poinsson brackets between the basis variables and $H$.
We can classify all the dynamical variables into two categories, $\{ O \} = \{ O_{e}\} \cup \{ O_{o} \}$. They satisfy $\langle \{ O_{e}, O_{e}^{\prime} \} \rangle =0$ and $\langle \{ O_{o}, O_{o}^{\prime}\} \rangle =0$.
A natural classification strategy that fulfils this requirement is
\begin{eqnarray}    \label{eq.36}
&&  \{O_{e}\} = \{ f(q) \prod_i p_i^{m_i} \,\, \Big{|} \,\, \sum_i m_i=2k, \,\, k \in \mathbb{Z} \}, \nonumber \\
&& \{O_{o}\} = \{ g(q) \prod_i p_i^{n_i} \,\, \Big{|} \,\,\sum_i n_i=2k+1, \,\, k \in \mathbb{Z}\}.\nonumber \\
&&
\end{eqnarray}
$f(q)$ and $g(q)$ are arbitrary functions of $q$.
If $H$ has the form $H = \sum_i p_i^2/(2\mu_i) + V(q)$, it is easy to prove that $ \{ \{ O_e, H \}, H \}  \in \{O_e\}$ and $ \{ \{ O_o, H \}, H \}  \in \{O_o\}$. That is, a basis inside $\{O_e\}$ or $\{O_o\}$ will remain so after being acted twice by $\{..., H\}$. We therefore consider to iterate the EOM twice and truncate the high order variable $\{ \{\vec{A}, H \}, H \}$. Choosing the basis operators $\{A_i\}$ from one of the subspaces, we have $\langle \{ \vec{A}, \vec{A}^{\dag} \} \rangle =0$. The second-order EOMs in matrix form are then obtained as
\begin{eqnarray}    \label{eq.37}
&&    \omega^2 G( \vec{A} \big| \vec{A}^{\dag} )_{\omega} =  -\langle \{\{ \vec{A}, H \} , \vec{A}^{\dag}\} \rangle - G( \{ \{\vec{A}, H \}, H \} \big| \vec{A}^{\dag} )_{\omega}, \nonumber \\
&&
\end{eqnarray}
and
\begin{eqnarray}    \label{eq.37.5}
&&    \omega^2 G( \vec{A} \big| \vec{A}^{\dag} )_{\omega} = \langle \{ \vec{A} , \{ \vec{A}^{\dag}, H \} \} \rangle - G( \vec{A} \big| \{ \{\vec{A}^{\dag}, H \}, H \} )_{\omega}. \nonumber \\
&&
\end{eqnarray}

To make PTA, we define the inner product of two variables $A$ and $B$ as
\begin{equation}    \label{eq.36.5}
 (A|B) \equiv  \langle \{ A^{\ast}, \{B, H\} \} \rangle.
\end{equation}
Considering that $\{ \{ \vec{A}, H\}, H \}$ contains only nonzero-frequency component (see Appendix B), we approximate it as 
\begin{equation}    \label{eq.38}
\{ \{ \vec{A}, H\}, H \} \approx - {\bf M}^{T} \vec{\bar{A}},
\end{equation}
where $\bar{A}_i \equiv A_i - A_{i0}$ is the nonzero-frequency component of $A_i$. This approximation is an extension of the Tyablikov-type decoupling approximation\cite{Tyablikov1} $G(OA|B)_{\omega} \approx \langle O \rangle G(A|B)_{\omega}$ to multicomponent case. It decouples the hierarchical EOMs into closed linear equations of GFs that are easy to solve. In the present work, as for the quantum systems\cite{Fan1}, we determine the expansion coefficients ${\bf M}$ by projection. 
Projecting Eq.(\ref{eq.38}) to $A_k$ and using the properties of static component $A_{i0}$ listed in Appendix B, we obtain ${\bf M} = {\bf I }^{-1}{\bf L}$.
Here, the Liouville matrix ${\bf L}$ is defined as
\begin{equation}    \label{eq.39}
  L_{ij} = -( A_i | \{ \{A_j, H \}, H \} ).
\end{equation}
${\bf I}$ is the inner product matrix with elements ${\bf I}_{ij}=(A_i|A_j)$. Using Eqs.(\ref{eq.5.13.5}) and (\ref{eq.8.3}), we find
$I_{ij} = \beta \langle \{A_i^{\ast}, H \} \{ A_j, H \} \rangle$ and $L_{ij} = \beta \langle \{ \{A_i^{\ast}, H \}, H\} \{ \{ A_j, H \}, H\} \rangle$. Both ${\bf I}$ and ${\bf L}$ are thus positive semidefinite Hermitian matrices. ${\bf M}$ is then guaranteed to have real positive eigen values.

The above method for determining ${\bf M}$ in Eq.(\ref{eq.38}) has several advantages over the traditional decoupling methods\cite{Tyablikov1}. It gives the best linear approximation of $\{ \{\vec{A}, H\}, H\}$ in the subspace $\{ A_i \}$ in the sense that the distance between $\{ \{ \vec{A}, H\}, H \}$ and $ -{\bf M}^{T} \vec{\bar{A}} $ (defined with respect to the given inner product) is the minimum one among all choices of ${\bf M}$. It fulfils the physical requirements that a GF has only real simple poles and that $G(A_i|A_i)_{\omega}$ has positive weights in $\omega > 0$ regime. When the basis $\{ A_i \}$ is complete, or it contains certain subspace of eigen modes of $H$, this approximation becomes exact. So we expect that PTA is a good approximation when the coordinates of the eigen modes of $H$ are adequately expressed by the linear combination of basis variables. Study shows that as the basis is enlarged, the approximation is improved systematically\cite{TM40.5}.

Substituting Eq.({\ref{eq.38}}) into Eq.(\ref{eq.37}), an approximate solution of GF matrix is obtained as
\begin{equation}    \label{eq.40}
G( \vec{A}\big| \vec{A^{\dag}} )_{\omega} \approx \left( \omega^2 - {\bf M}^{T} \right)^{-1} {\bf I}^{T},
\end{equation}
or in terms of ${\bf U}$ and ${\bf \Lambda}$,
\begin{eqnarray}    \label{eq.42}
   G( \vec{A}\big| \vec{A^{\dag}} )_{\omega} &\approx & \left( {\bf IU} \right)^{\ast} \left( \omega^2 {\bf 1} - {\bf \Lambda} \right)^{-1} \left({\bf IU} \right)^{T} .
\end{eqnarray}
Here, ${\bf U}$ is the eigenvector matrix of ${\bf M}$ and ${\bf \Lambda} = diag(\lambda_1, \lambda_2, ..., \lambda_n)$ is the eigenvalue matrix, with $\lambda_k$ being real and $\lambda_k \geqslant 0$ for all $k$. They can be obtained by solving the generalized eigen-value problem,
\begin{equation}    \label{eq.43}
   {\bf LU} = {\bf IU \Lambda}.
\end{equation}
${\bf U}$ satisfies the generalized unitary condition ${\bf U}^{\dagger} {\bf I} {\bf U} = {\bf 1}$.
The element of GF matrix reads
\begin{equation}    \label{eq.44}
G( A_i | A_j^{\ast} )_{\omega} \approx \sum_{k} \frac{\left( {\bf IU} \right)^{\ast}_{ik} \left( {\bf IU} \right)_{jk}}{\omega^2 - \lambda_k}.
\end{equation}
The fluctuation-dissipation theorem Eq.(\ref{eq.20.4}) produces the following equations for the averages, 
\begin{equation}    \label{eq.45}
   \langle A_j^{\ast} A_{i}  \rangle   \approx \sum_{k} \frac{ \left( {\bf IU} \right)^{\ast}_{ik} \left( {\bf IU}\right)_{jk}}{\beta \lambda_k} +  \langle A_{j0}^{\ast} A_{i0}  \rangle.
\end{equation}
An equivalent expression is
\begin{equation}    \label{eq.46}
   {\bf C} \approx  \frac{1}{\beta} {\bf I}{\bf L}^{-1} {\bf I} + {\bf C}_0,
\end{equation}    
which does not require the solution of a generalized eigen-value problem. Here, ${\bf C}$ and ${\bf C}_0$ are the correlation matrices defined below Eq.(\ref{eq.20.4}).   

Similarly, we can derive expressions for the GF $G( \vec{A}\big|O^{\ast} )_{\omega}$ for arbitrary variable $O$ as
\begin{eqnarray}    \label{eq.47}
G( \vec{A}\big|O^{\ast} )_{\omega} &\approx & \left( \omega^2 {\bf 1}- {\bf M}^{T} \right)^{-1} ( O\big| \vec{A} ) \nonumber \\
    &= & \left( {\bf IU} \right)^{\ast} \left( \omega^2 {\bf 1} - {\bf \Lambda} \right)^{-1} {\bf U}^{T}( O\big|\vec{A} ).
\end{eqnarray}
The averages are given as
\begin{equation}    \label{eq.48}
   \langle O^{\ast} A_{i}  \rangle  \approx \sum_{k,p} \frac{\left( {\bf IU}\right)^{\ast}_{ik} {\bf U}_{pk} (O|A_{p}) }{\beta \lambda_k} + \langle O^{\ast}_{0} A_{i0}  \rangle,
\end{equation}
or in the vector form,
\begin{equation}    \label{eq.49}
   \langle O^{\ast} \vec{A}^{T} \rangle  \approx \frac{1}{\beta} (O\big|\vec{A}^{T}) {\bf L}^{-1}{\bf I} + \langle O^{\ast}_{0} \vec{A_{0}}^{T} \rangle.
\end{equation}

If the matrices ${\bf L}$ and ${\bf I}$ are expressible by ${\bf C}$, i.e., ${\bf L = L(C) }$ and ${\bf I = I(C) }$, and if ${\bf C}_0$ can be calculated, Eq.(\ref{eq.45}) (or Eq.(\ref{eq.46})) closes the equation for ${\bf C}$. Solving this equation can provide approximate values for the static correlation functions. The GF is then obtained from Eq.(\ref{eq.40}). If ${\bf L}$ and ${\bf I}$ involve the averages other than elements of ${\bf C}$, one needs to resort to Eq.(\ref{eq.49}) for additional equations. The conservation relation Eq.(\ref{eq.8.1}) could also provide additional constraints on the involved averages. The whole scheme is similar to the quantum case \cite{Fan1}.

\end{section}

\begin{section}{The one-dimensional nonlinear $\phi^4$ lattice}

In this section, we apply PTA to the classical one-dimensional $\phi^4$ lattice model with the following Hamiltonian,

\begin{equation}      \label{eq.49.1}
  H = \sum_{i=1}^L \left[ \frac{p_i^2}{2m} + V(x_i-x_{i-1}) + U(x_i) \right],
\end{equation}
with
\begin{eqnarray}      \label{eq.49.2}
  && V(x_i-x_{i-1}) = \frac{K}{2} (x_i-x_{i-1})^2, \nonumber \\
  && U(x_i) = \frac{\gamma}{4} x_i^4.
\end{eqnarray}
Here, $L$ is the total number of classical particles. $x_i$ represents the deviation of the $i$-th particle from its equilibrium position. The position of the $i$-th particle is $q_i = ia + x_i$. $K$ is the nearest-neighbor coupling strength and $\gamma$ the coefficient of the on-site potential. Here, we use periodic boundary condition and set the lattice constant $a = 1$ and mass $m=1$.  This model can be obtained by discretizing the classical $\phi^4$ field theory\cite{DB31,KA32}. It describes a harmonic-coupled chain of particles, with each particle in a local quartic potential well.

To employ the translational symmetry of $H$, we express Eq.($\ref{eq.49.1}$) in wave vector space as
\begin{eqnarray}      \label{eq.49.4}
&& H = \sum_{k} H_{k},  \nonumber \\
&& H_{k} =  \frac{ P_{k}P_{k}^{\ast}}{2} + \frac{\omega_0(k)^2}{2}Q_{k}Q_{k}^{\ast} + \frac{\gamma}{4}Q_{k}R_{k}^{\ast}.
\end{eqnarray}
Here, $\omega_0(k)^2=2K[1-\cos(k)]$, $Q_{k}=1/\sqrt{L}\sum_j e^{-ijk}x_j$, and $R_{k}=1/\sqrt{L} \sum_j e^{-ijk}x_j^3$.
The conjugate momentum of $Q_{k}$ is $P_{k}=1/ \sqrt{L}\sum_j e^{ijk}p_j$. They satisfy the relation $\{Q_{k}, P_{k'} \} = \delta_{k, k'}$.

Similar to the Fermi-Pasta-Ulam (FPU)-$\beta$ model\cite{KA39}, this $\phi^4$ lattice model has an interesting scaling property\cite{KA32}. Using the scaling transformation
\begin{equation}      \label{eq.49.3}
 p_i=\frac{K}{\sqrt{\gamma}}\tilde{p}_i, \hspace{2em} x_i=\sqrt{\frac{K}{\gamma}}\tilde{x}_i,
\end{equation}
one obtains
\begin{equation}      \label{eq.49.3.1}
 H=\frac{K^2}{\gamma}\widetilde{H},
\end{equation}
where the dimensionless Hamiltonian $\widetilde{H}$ reads
\begin{equation}      \label{eq.49.3.2}
 \widetilde{H} = \sum_{i=1}^L \big [ \frac{\tilde{p}_i^2}{2} + \frac{1}{2} (\tilde{x}_i-\tilde{x}_{i-1})^2 + \frac{1}{4}\tilde{x}_i^4 \big ].
\end{equation}
This implies a scaling form of physical quantities. For examples,
\begin{eqnarray}      \label{eq.49.3.4}
&& \langle x_i^2 \rangle (K, \gamma, T) = \frac{K}{\gamma} \langle x_i^2 \rangle \left(1, 1, \frac{\gamma T}{K^{2}} \right),  \nonumber \\
&& \langle x_i^4 \rangle (K, \gamma, T) = \frac{K^2}{\gamma^2} \langle x_i^4 \rangle \left( 1, 1, \frac{\gamma T}{K^{2}} \right), \nonumber \\
&& C_v(K, \gamma, T) = C_v \left(1, 1, \frac{\gamma T}{K^{2}} \right), \nonumber \\
&& \xi(K, \gamma, T))= \xi \left(1, 1, \frac{\gamma T}{K^{2}} \right).
\end{eqnarray}
Here, $C_v= (1/L) \partial \langle H\rangle / \partial T$ is the isovolumetric specific heat capacity and $\xi$ is the correlation length defined as $\langle x_i x_j\rangle \sim  e^{-|i-j|/\xi}$, $|i-j| \rightarrow \infty$.

Similar scaling relations exist for GFs. By moment expansion of GF, we obtain
\begin{eqnarray}    \label{eq.49.3.5}
&&  G(x^m p^n | x^r p^s)(K, \gamma, T, \omega)  \nonumber \\
&=& K^{\theta_K} \gamma^{\theta_{\gamma}} G(x^m p^n | x^r p^s)\left(1, 1, \frac{\gamma T}{K^{2}}, \frac{\omega}{\sqrt{K}} \right),
\end{eqnarray}
with the scaling exponents $\theta_K = (m+r)/2+n+s-2$ and $\theta_\gamma = -(m+r+n+s)/2+1$. In the above equation, $x^{m}$ is the short-hand notation for $x_1^{m_1} x_2^{m_2} ... x_L^{m_L}$ with $m_1 + m_2 + ... + m_L = m$. $p^{n}$ is a similar notation. Taking $m=r=1$ and $n=s=0$, we obtain
\begin{equation}   \label{eq.49.3.6}
  G(Q_{k}|Q_{k}^{\ast})(K, \gamma, T, \omega) = \frac{1}{K} G(Q_{k}|Q_{k}^{\ast}) \left( 1, 1,\frac{\gamma T}{K^{2}}, \frac{\omega}{\sqrt{K}} \right).
\end{equation}
Using the spectral decomposition of classical GF\cite{Cavallo2} $G(\omega) = \sum_{k} W_k/(\omega - E_k)$ and assuming that poles $E_k$'s and weights $W_k$'s scale independently, we find the following scaling relations,
\begin{eqnarray}   \label{eq.49.3.7}
 &&   E_k(K, \gamma, T) = \sqrt{K} E_k(1, 1, \frac{\gamma T}{K^{2}}), \nonumber \\
  &&  W_k(K, \gamma, T) = K^{\theta_K} \gamma^{\theta_{\gamma}} W_k\left(1, 1, \frac{\gamma T}{K^{2}} \right),
\end{eqnarray}
with $\theta_K =(m+r)/2+n+s-3/2$ and  $\theta_\gamma = -(m+r+n+s)/2+1$.
Here, we have allowed the quasi-particle energy $E_k$ to be temperature dependent. In particular, Eq.(\ref{eq.49.3.7}) implies that the phonon gap has the scaling relation $\Delta(K, \gamma, T) = \sqrt{K} \Delta(1, 1, \gamma T/K^2)$.
Since the projection truncation of GFs conform to the scaling transformation, we expect that PTA obeys all the above scaling relations. Indeed, with PTA numerical data, we numerically checked Eqs.(\ref{eq.49.3.4}), (\ref{eq.49.3.6}), and that for $\Delta(K, \gamma, T)$ and found perfect agreement.

Some exact relations about $H$ can be obtained from the conservation relation Eq.(\ref{eq.8.1}). Taking $O = x_i^np_i$ and $O = Q_kP_k$ in Eq.(\ref{eq.8.1}), respectively, we obtain
\begin{equation}    \label{eq.49.4.1}
\gamma \langle x_i^{n+3} \rangle + 2K\langle x_i^{n+1} \rangle -
K \left( \langle x_i^{n}x_{i-1} \rangle + \langle x_i^{n}x_{i+1} \rangle \right) = nT\langle x_i^{n-1} \rangle
\end{equation}
and
\begin{equation}    \label{eq.49.4.2}
\omega_0(k)^2 \langle Q_kQ_k^{\ast} \rangle + \gamma \langle  Q_kR_k^{\ast} \rangle = T.
\end{equation}
Eq.$\eqref{eq.49.4.1}$ with $n = 1$ and Eq.$\eqref{eq.49.4.2}$ are the generalized equipartition theorem (GET) in real space and wave vector space, respectively. We have numerically checked that the PTA results, both from B1 and B2 bases (to be defined below) fulfil the above GET.

In the high temperature limit, due to large amplitude of oscillations, the $x_i^4$-term in $H$ dominates the energy and nonlocal correlation between particles can be ignored\cite{NL38}. One expects that the system be well described by an independent anharmonic oscillator with Hamiltonian $H_s = p^2/2 + (\gamma/4) x^4$. It gives, in the limit $T \rightarrow \infty$,
\begin{equation}    \label{eq.49.4.3}
\langle x^{2n}\rangle=2^n\frac{\Gamma(\frac{n}{2}+\frac{1}{4} )}{\Gamma( \frac{1}{4} )}\gamma^{-\frac{n}{2} }T^{ \frac{n}{2} }.
\end{equation}
Here, $n=0,1,2, ...$. $\Gamma(1/4)$ and $\Gamma(n/2+1/4)$ are complete $\Gamma$ functions.
Through integral of equation of motion, the kinetic-temperature dependence of the frequency of this single oscillator is obtained exactly as
\begin{equation} \label{Eq.w0}
 \omega_{\rm single}(T)=\frac{\sqrt{2\pi}\Gamma(\frac34)3^{\frac14}}{\Gamma(\frac14)}T^{\frac14}\gamma^{\frac14} \approx 1.115 T^{\frac14}\gamma^{\frac14}.
\end{equation}

\end{section}

\begin{section}{Applying PTA to the one-dimensional $\phi^4$ lattice model}

\begin{subsection}{Formalism}

To apply PTA to the one-dimensional $\phi^4$ lattice model Eq.\eqref{eq.49.4}, in this work we consider the following two bases.
(1) basis B1: $\vec{A}_1 = (Q_{k})^{T}$; and (2) basis B2: $\vec{A}_2=(Q_{k}, R_{k})^{T}$.
$Q_{k}$ and $R_{k}$, as defined below Eq.(\ref{eq.49.4}), are Fourier transformations of $x_i$ and $x^{3}_i$, respectively. Considering that the Hamiltonian is even under the two global parity transformations $x_i \rightarrow -x_i$ and $p_i \rightarrow -p_i$ ($i=1,2,...,L$), any conserved quantity $X(\{x_i, p_i\})$ should also be even, given that $\sum_i [\partial X/\partial x_i\partial H/\partial p_i ] = \sum_i [\partial X/\partial p_i\partial H/\partial x_i ]$. Since $Q_{k}$ and $R_{k}$ are odd under parity transformation, they do not have zero-frequency components, i.e., $Q_{k0} = R_{k0}=0$. Therefore, we use ${\bf C}_0 = 0$ in Eqs.(\ref{eq.20.4}) and (\ref{eq.46}).  
As will be seen below, PTA under basis B1 gives identical results to self-consistent phonon theory (i.e., the quadratic variational method)\cite{JL40}. PTA with basis B2 gives improved results over B1.

\begin{subsubsection}{Basis B1: $\vec{A}_1 = \left(Q_{k} \right)^{T}$}

For this one-dimensional basis, we obtain
\begin{eqnarray}      \label{eq.49.5}
&& I_{k,k'} = \delta_{k,k'}, \nonumber \\
&& L_{k,k'} = \omega(k)^2 \delta_{k,k'} .
\end{eqnarray}
Here, $\omega(k)^2 = \omega_0(k)^2 + (3\gamma)/L \sum_{k'}\langle Q_{k'} Q_{k'}^{\ast}\rangle$.
Employing the spectral theorem and noting $Q_{k0} = 0$ for the $\phi^4$ model, we get the self-consistent equation
\begin{eqnarray}      \label{eq.49.6}
 && \langle Q_{k} Q_{k}^{\ast}\rangle = \frac{1}{\beta \omega(k)^2}, 
\end{eqnarray}
From the pole of $G(Q_k|Q_k^{\ast})_{\omega}$, we obtain the phonon dispersion $\omega(k)=\sqrt{\omega_0(k)^2 + 3\gamma \langle x_i^2 \rangle}$.
It is same as the result from self-consistent phonon theory (i.e. quadratic variational method)\cite{JL40}.
In the PTA study of the spinless fermion model\cite{Ma1}, the single anharmonic oscillator model\cite{TM40.5}, and the FPU-$\beta$ model, we all found that under one-dimensional single particle basis, PTA results are identical to those from the variational method with a quadratic reference Hamiltonian. Though not yet proved rigorously, we believe that this is true in general. With enlarged basis size, PTA within GF EOM method could provide a convenient scheme for systematically going beyond the traditional variational approximation.

\end{subsubsection}

\begin{subsubsection}{Basis B2: $\vec{A}_2 = \left(Q_{k}, R_{k}\right)^{T}$}

For this two-dimensional basis, we obtain ${\bf I}_{kk^{\prime}} = {\bf I}_{k} \delta_{k, k^{\prime}}$, ${\bf L}_{kk^{\prime}} = {\bf L}_{k} \delta_{k, k^{\prime}}$, with
\begin{equation}    \label{eq.50}
  {\bf I}_{k} = \left(
\begin{array}{cc}
1 &  3f_1  \\
3f_1 &  9f_2  \\
\end{array}
\right),
\end{equation}
and
\begin{equation}    \label{eq.51}
 {\bf L}_k = \left(
\begin{array}{cc}
\omega(k)^2 &  3\omega_0(k)^2 f_1+9\gamma f_2 \\
3\omega_0(k)^2 f_1+9\gamma f_2 & L_{22} \\
\end{array}
\right).
\end{equation}
Here,
\begin{equation}    \label{eq.51.5}
 L_{22} =  \frac{54}{\beta}f_1+54Kf_2+45\gamma f_3 - 36Kf_4 -18K \cos(k)f_5.
\end{equation}
The real functions $f_1 \sim f_5$ are defined as
\begin{eqnarray}    \label{eq.52}
 && f_{1} = \frac{1}{L} \sum_{k_1}\langle Q_{k_1}Q_{k_1}^{\ast}\rangle,  \nonumber \\
 && f_{2} = \frac{1}{L} \sum_{k_1}\langle Q_{k_1}R_{k_1}^{\ast}\rangle,  \nonumber \\
 && f_{3} = \frac{1}{L} \sum_{k_1}\langle R_{k_1}R_{k_1}^{\ast}\rangle,  \nonumber \\
 && f_{4} = \frac{1}{L} \sum_{k_1} \cos(k_1)\langle Q_{k_1}R_{k_1}^{\ast}\rangle, \nonumber \\
 && f_{5} = \frac{1}{L} \sum_{k_1}e^{-ik_1}\langle Q_{k_1}^{\ast}O_{k_1}\rangle.
\end{eqnarray}
The variable $O_{k}$ in $f_{5}$ in the above equation is defined as $O_{k}=(1/\sqrt{L})\sum_j e^{-ijk}x_j^2 x_{j+1}$. The average $\langle Q_{k_1}^{\ast}O_{k_1}\rangle$ needs to be calculated from the new GF $G(\vec{A}_2 \big| O_{k}^{\ast})_{\omega}$, following Eq.(\ref{eq.47}). The inner products used in this process are
\begin{eqnarray}    \label{eq.52.1}
 && (O_{k}|Q_{k})=e^{-ik}f_1 + \frac{2}{L} \sum_{k_1} \cos(k_1)\langle Q_{k_1}Q_{k_1}^{\ast}\rangle,  \nonumber \\
 &&  (O_{k}|R_{k})=6f_4 + 3e^{-ik}f_5 .
\end{eqnarray}
In the derivation of above equations, we have used the exact relation $\langle p^2 \rangle = 1/\beta$. The positive-definiteness of ${\bf I}$ amounts to $\langle  x^4 \rangle - \langle x^2 \rangle^2 >0$, a physical requirement. The positive-definiteness of ${\bf L}$ also represents constraints on the averages.

\end{subsubsection}

\end{subsection}

\begin{subsection}{Numerical Results }

Below, we present the numerical results obtained by solving the self-consistent equation Eq.(\ref{eq.46}) for the above two bases. Due to the scaling properties Eqs.$(\ref{eq.49.3}) \sim (\ref{eq.49.3.2})$, unless otherwise specified, we choose the model parameters $K = \gamma = 1$.

\begin{subsubsection}{$\phi^4$ lattice}
\begin{figure}
 \vspace*{-0.0cm}
\begin{center}
  \includegraphics[width=260pt, height=200pt, angle=0]{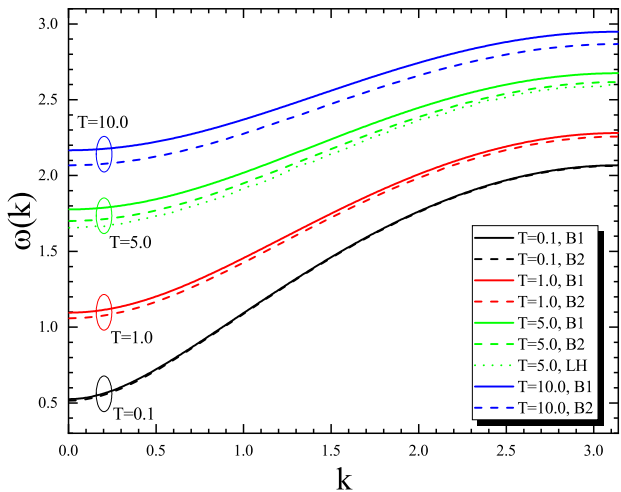}
  \vspace*{-1.0cm}
\end{center}
  \caption{(color online) Dispersion relation $\omega(k)$ at different temperatures. The solid and dashed lines are results from B1 and B2 bases, respectively. The green dotted line for $T=5.0$ (LH) is from the lower bound harmonic variation of free energy obtained by Liu {\it et al.}\cite{JL40}. 
}   \label{Fig1}
\end{figure}

Basis B1 produces a single excitation at $\omega(k)$. For basis B2,  $G(Q_k | Q_k^{\ast})_{\omega}$ has two poles on the positive frequency axis. One of them carries most of the weight. It is regarded as the phonon excitation. The other pole has only tiny weight and is located at noninteger times of the phonon frequency. It is more like a satellite peak of the main peak rather than the overtone or combination tone frequently observed in molecular/crystal vibrational spectrum. 
This satellite peak is dependent on the arbitrary choice of the basis is not a physical effect but rather a non-desired parasite.
The appearance of combination tone requires multiple fundamental frequencies, which is not the case for the monoatonic $\Phi^{4}$ lattice model studied here. In classical system, anharmonicity makes the fundamental frequency energy dependent and thus gives broadened peak in the spectral function at finite $T$\cite{YO1}. The overtone of the $\Phi^{4}$ lattice may manifest itself in the spectral function as a weak broad peak at integer times of the phonon frequency, similar to what we observed in molecular dynamics(MD) simulation of the power spectrum (not shown). It is expected to obtain only in the large basis limit of PTA. Obtained from the two-dimensional basis B2, the pole with the tiny weight may well be a precursor of the broadening of the main peak. Detailed investigation of this issue using larger basis will be the subject of later study.

In Fig.\ref{Fig1}, we show phonon dispersions obtained from B1 and B2 bases for a series of temperatures. For a given temperature, $\omega(k)$ is a monotonously increasing function with a gap at $k=0$. B1 and B2 only produces quantitative differences. The difference approaches zero at low temperature (say $T = 0.1$) and enlarges with increasing temperature. The is expected since at low $T$, the anharmonic potential barely influences the small amplitude oscillation of particles and the quadratic variational approximation is adequate. At higher $T$, the anharmonic effect taken into account by B2 becomes more significant. For a given $T$, B2 basis always produces lower $\omega(k)$ than B1 does, similar to the results of Ref.\cite{JL40} where $\omega(k)$ from the lower bound harmonic variation lies below that from the upper bound variation. The former one is expected to be more accurate since correlations are treated more adequately.

Note that the damping of the phonon excitation due to the $x_{i}^{4}$-term is not described within B2 basis. As temperature increases, the phonon peak in the spectral function should be broadened significantly and finally smeared in the high temperature limit (e.g. $T = 10$) where phonon is no longer well defined\cite{YO1}. In contrast, the B2 basis produces two $\delta$-peaks in the spectral function at finite temperature. The quantitative improvement in the static quantities by B2 basis therefore mainly comes from better description of the spectral moments but not the damping. To describe the damping effect that is indispensable for the heat conductivity study, we need to either use a much larger basis dimension in PTA, or supplement PTA with a memory function calculation.

\begin{figure}
 \vspace*{-0.0cm}
\begin{center}
  \includegraphics[width=260pt, height=200pt, angle=0]{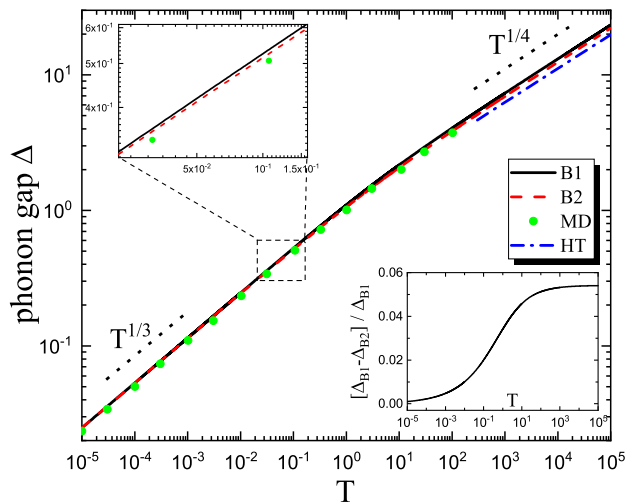}
  \vspace*{-1.0cm}
\end{center}
  \caption{(color online) Phonon gap as a function of temperature $T$. Green solid dots are from MD simulation. The blue dash-dotted line in high temperature regime (HT) is for Eq.(\ref{Eq.w0}). The black dotted lines are for guiding eyes. Upper-left inset: enlarged part of main figure. Lower-right inset: relative difference between B1 and B2 results.
}   \label{Fig2}
\end{figure}

At finite temperature, a gap $\Delta(T)$ emerges in the phonon spectrum at $k=0$ due to the existence of on-site potential. This gap will significantly affect the heat transport behavior of the $\phi^4$ model. Fig.$\ref{Fig2}$ compares $\Delta(T)$ obtained from various methods. B1 and B2 give qualitatively similar $\Delta(T)$. It has the low- and high-temperature asymptotic power laws as $\Delta(T) \sim T^{1/3}$ ($T \rightarrow 0$) and $\sim T^{1/4}$ ($T \rightarrow \infty$). The lower-right inset shows the relative error between B1 and B2 results. It increases from zero at $T=0$ and saturates in the high $T$ limit.

The low-/high-temperature comparison of $\Delta(T)$ deserves separate discussions. In the high temperature limit, the inter-particle couplings are negligible and particles move basically independently. 
Then Eq.(\ref{Eq.w0}) gives the exact gap $\Delta(T)$ in infinite temperature limit and an upper bound of gap at finite temperature.  
In Fig.~\ref{Fig2}, $\omega_{single}(T)$ (see Eq.\eqref{Eq.w0}) is plotted as a blue dash-dotted line in the high-temperature regime. 
We see that both B1 and B2 are only slightly greater than $\omega_{single}(T)$, and more importantly, B2 is in even better agreement.

In low temperature regime, $\langle x_i^2\rangle \propto T^{2/3}$\cite{DB31}, and thus the lattice can no longer be regarded as independent particles. Eq.~(\ref{Eq.w0}) does not apply.
In such a case, in order to evaluate B1 and B2 we turn to calculating $\Delta(T)$ numerically by MD simulations. 
The simulations are carried out in a lattice with periodic boundary condition and $L=1000$ particles.
A set of randomly chosen initial states are extracted from the microcanonical ensemble with
fixed energy density $\langle E\rangle$ which corresponds to the desired temperature $T$.
The power spectrum $S_k(\omega)$ of the mode with wave vector $k$,
i.e., the Fourier transform of $P_k$\cite{PhysRevLett.120.144301} is then calculated after long enough transient time. The profile of each power spectrum is basically a single peak.
The frequency $\omega(k)$ can be simply determined by the location of the peak.
The so-measured $\Delta \equiv \omega(0)$ as a function of $T$ is plotted in Fig.~\ref{Fig2} as well as in the upper-left inset as green circles.
Again, we see that both B1 and B2 agree with the numerical simulation very well and the agreement of B2 is even better.
In summary, B1 and B2 bases give correct low- and high-temperature exponents of $\Delta(T)$. B2 gives quantitatively improved results over B1.

\begin{figure}
 \vspace*{0.250cm}
\begin{center}
  \includegraphics[width=250pt, height=200pt, angle=0]{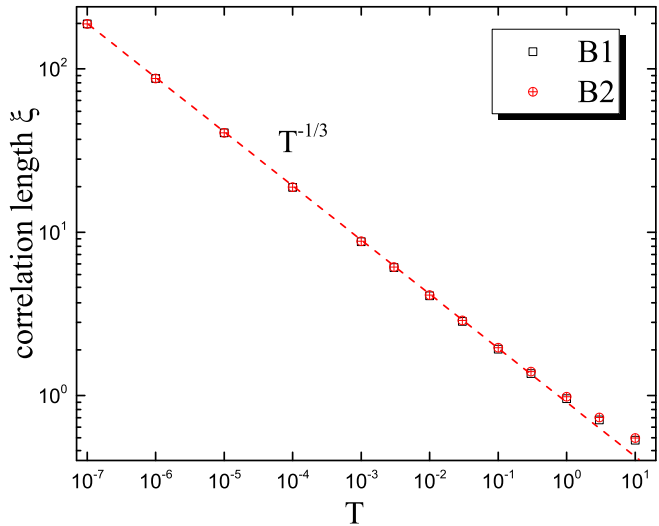}
  \vspace*{-1.0cm}
\end{center}
  \caption{(color online) Correlation length $\xi$ as a function of temperature $T$. The red dashed line marks the $T^{-1/3}$ power law. 
}   \label{Fig3}
\end{figure}

Figure $\ref{Fig3}$ shows the temperature dependence of correlation length $\xi$, defined by $\langle x_i x_j\rangle \sim e^{-|i-j|/\xi}$ ($|i-j| \gg 1$). Both bases give $\xi \propto T^{-1/3}$ in the low temperature limit. According to Eq.(\ref{eq.49.3.4}), the limit $T \to 0$ is equivalent to $\gamma \to 0$ or $K \to \infty$ for $\xi$. In this limit, the particles move along the chain with unbound but locked-in displacements $x_i = x_j$, giving $\xi = \infty$.  Further, we find that the numerical results fulfil $\langle x_i x_j\rangle = \langle x_i^2 \rangle e^{-|i-j|/\xi}$ not only for large $|i-j|$ but also for short range. This helps understanding the power exponent $1/3$. Assigning $j=i+1$ in the above equation, we have $\xi = 1/ ln \left[ \langle x_i^2 \rangle/\langle x_i x_{i+1}\rangle \right]$. Employing the exact low temperature asymptotic behaviors $\langle x_i^2 \rangle \approx \langle x_i x_{i+1}\rangle \propto T^{2/3}$ and $\langle x_i^2 \rangle - \langle x_i x_{i+1}\rangle \propto T$ (see Fig.$\ref{Fig5}$ and its discussion), we obtain the relation $\xi \propto T^{-1/3}$.
In high temperatures (say $T \sim 1$), both B1 and B2 results deviate from $T^{-1/3}$ law, signalling that the system enters a strong chaotic regime\cite{JL40,MP42,MP43}.

\begin{figure}
 \vspace*{-1.0cm}
\begin{center}
  \includegraphics[width=350pt, height=280pt, angle=0]{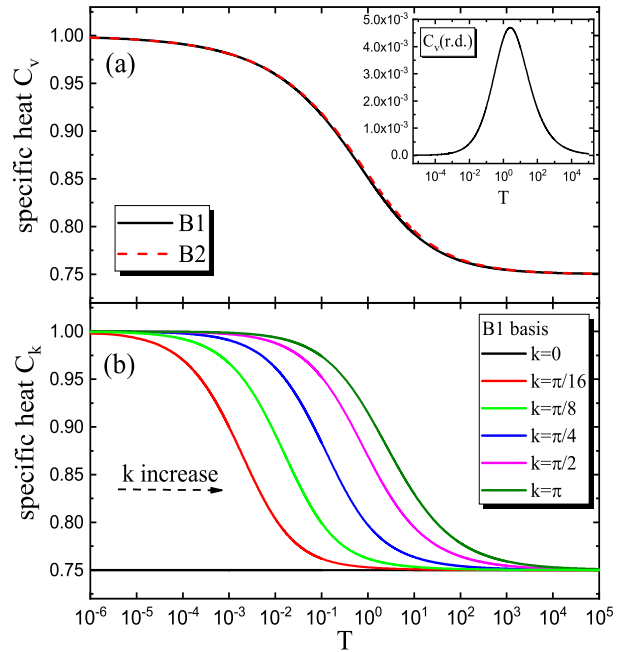}
  \vspace*{-1.0cm}
\end{center}
  \caption{(color online) Specific heat. (a) $C_v$, and (b) $C_k$ as functions of temperature $T$. Inset of (a): The relative difference between specific heat $C_v$ from B1 and B2. $C_v(r.d)=\left[C_v(B2)-C_v(B1)\right]/C_v(B1)$.  
}   \label{Fig4}
\end{figure}

Figure $\ref{Fig4}$(a) shows the evolution of isovolumetric specific heat capacity $C_v$ with temperature. Again, the results from B1 and B2 bases have only slight difference.
With increasing temperature, $C_v$ gradually decreases from $C_v=1.0$ at $T=0$ (equivalent to the harmonic limit) to $3/4$ in the high-temperature limit.
This behavior is similar to that of the FPU-$\beta$ model\cite{DH41}.
It is not a coincidence, but due to the thermodynamic similarity between the $\phi^4$ model and the FPU-$\beta$ model in the high- and low- temperature limits.
By analysis of GET, we obtain $C_v = 1.0 - (\gamma/4) \partial \langle x_i^4\rangle / \partial T$ and the low/high temperature asymptotic behaviors $\langle x_i^4\rangle \ll T$ ($T\to 0$) and $\langle x_i^4\rangle \approx T$ ($T\to \infty$). We thus confirm that the results of $C_v$ in Fig. $\ref{Fig4}$(a) are also exact in the high- and low- temperature limits.
The crossover of $C_v$ from low to high temperature occurs at around $T=1.0$. As shown in the inset of Fig.$\ref{Fig4}$(a), the relative difference between B1 and B2 results is less than $0.5\%$, with the maximum near the crossover temperature.

Figure $\ref{Fig4}$(b) presents the temperature dependence of single-mode specific heat $C_k$, defined as $C_k=\partial\langle H_k\rangle /\partial T$ and $H_k$ defined in Eq.(\ref{eq.49.4}).
Only B1 result is shown here since the B2 one is quantitatively similar. $C_k(T)$ looks similar to $C_v(T)$, with a crossover temperature increasing with $k$. This is because the dispersion function $\omega(k)$ increases monotonously with $k$, as shown in Fig.$\ref{Fig1}$. The excitation of larger momentum phonon requires more energy, resulting in greater specific heat $C_k$. Similarly, the asymptotic high- and low- temperature limits of $C_k$ are captured exactly by PTA.
Note that $C_{k=0}(T)=3/4$ for all $T$ is an exact result, since the GET Eq.$\eqref{eq.49.4.2}$ gives $\gamma \langle Q_{k=0} R_{k=0}^{\ast} \rangle = T$ and $\langle H_{k=0} \rangle =3T/4$.

\begin{figure}
 \vspace*{-0.20cm}
\begin{center}
  \includegraphics[width=320pt, height=260pt, angle=0]{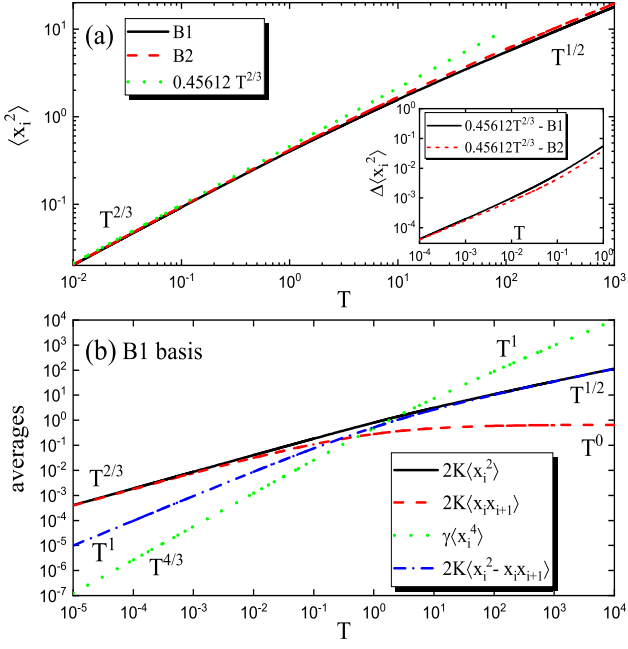}
  \vspace*{-1.0cm}
\end{center}
  \caption{(color online) Average of physical quantities as functions of $T$. (a) $\langle x_i^2 \rangle$ from B1 and B2 basis are compared. The dotted line shows $0.45612 T^{2/3}$ from the classical field method\cite{DB31}. Inset: differences between PTA and classical field results. (b) B1 basis results for $2K\langle x_i^2 \rangle$, $2K\langle x_i x_{i+1} \rangle$, $\gamma\langle x_i^4 \rangle$, and $2K(\langle x_i^2 \rangle-\langle x_i x_{i+1} \rangle)$ as functions of $T$.
  }   \label{Fig5}
\end{figure}
Figure $\ref{Fig5}$ shows the temperature dependence of several types of averages.
We focus on $\langle x_i^2 \rangle$ in Fig.$\ref{Fig5}$(a).
It is found that $\langle x_i^2 \rangle \propto T^{2/3}$ ($T \rightarrow 0$) and $\langle x_i^2 \rangle \propto T^{1/2}$ ($T \rightarrow \infty$), with a crossover temperature around unity. Our result at low temperature agrees quantitatively with that from the classical field method\cite{DB31} (dotted line in Fig.$\ref{Fig5}$(a)).
In the high temperature limit, $\langle x_i^2 \rangle \propto T^{1/2}$ can be understood from Eq.$\eqref{eq.49.4.3}$.
Quantitatively, $\langle x_i^2 \rangle$ obtained by the B2 basis is slightly larger than that of the B1 basis.
In the inset of Fig.$\ref{Fig5}$(a), we compare the two PTA results with $\langle x_i^{2} \rangle = 0.45612T^{2/3}$ of the classical field method. As expected, the B2 basis compares more favorable.
We also studied $\langle x_i^2 \rangle$ , $\langle x_i^4 \rangle$, and $\langle x_i x_{i-1} \rangle$, {\it etc.} using MD. The MD results (not shown) agree very well with B1 and B2 results.

For $\phi^4$ lattice model, Eq.(\ref{eq.49.4.1}) (with $n=1$) gives the GET $\gamma \langle x_i^4 \rangle + 2K \langle x_i^2 \rangle - 2K \langle x_i x_{i+1} \rangle = T$.
Figure $\ref{Fig5}$(b) shows the temperature dependence of all averages appearing in this equation, obtained from B1 basis.
Our numerical results satisfy the GET within numerical error.
All curves in Fig.$\ref{Fig5}$(b) have power law behavior in the low as well as high temperature limits, with distinct powers and similar crossover temperatures around unity. This value of crossover temperature is comparable to the strong stochasticity threshold temperature of the $\phi^4$ model\cite{JL40,MP42,MP43}.

Several noteworthy features of Figure $\ref{Fig5}$(b) are discussed in order. First, in $T \ll 1$, $\langle x_i^2 \rangle$ and $\langle x_i x_{i+1} \rangle$ have same leading power, while in $T \gg 1$, $\langle x_i x_{i+1} \rangle \ll \langle x_i^2 \rangle$. This is consistent with the temperature-dependent behavior of the correlation length shown in Fig.$\ref{Fig3}$, signalling the weakening of nonlocal correlation at high temperature.
Second, in the low temperature limit, $\langle x_i^2 \rangle \sim T^{2/3}$ and $\langle x_i^2 \rangle - \langle x_i x_{i+1} \rangle \sim T$, respectively. This is because when $T\to 0$ (equivalent to $\gamma \to 0$), the translational symmetry of the system gradually recovers. The independent dynamic variable is not $x_i$ but $x_{i+1} - x_i$. As a result, $\langle (x_i - x_{i+1} )^2\rangle = 2(\langle x_i^2 \rangle - \langle x_i x_{i+1}\rangle) \propto T$ according to the equipartition theorem.
Finally, for $\langle x_i^4 \rangle$, the variational approximation is considered to be reliable in $T \ll 1$. So $\langle x_i^4 \rangle \approx 3 \langle x_i^2 \rangle^2 \propto T^{4/3}$. In the high temperature limit, GET guarantees $\langle x_i^4 \rangle \sim T$.
In summary, the exact power exponents in the $T$-dependence of averages are obtained in Fig.$\ref{Fig5}$.

\begin{figure}
 \vspace*{-0.0cm}
\begin{center}
  \includegraphics[width=250pt, height=200pt, angle=0]{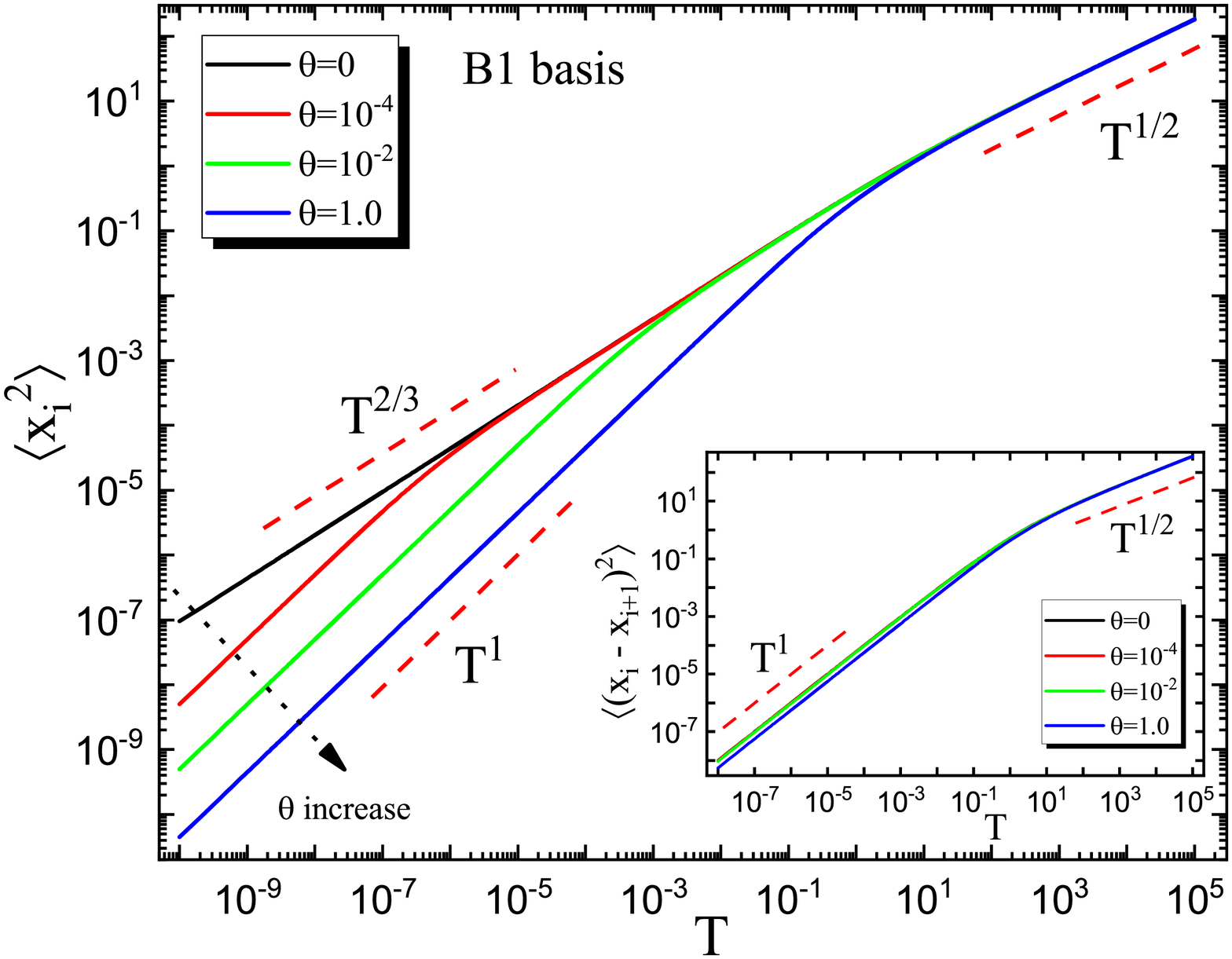}
  \vspace*{-1.0cm}
\end{center}
  \caption{(color online) $\langle x_i^2 \rangle$ as a function of $T$, for the modified $\phi^4$ model at various $\theta$ values. Inset: $\langle (x_i - x_{i+1})^2 \rangle$ as functions of $T$. The red dashed lines mark the corresponding power law.
}   \label{Fig6}
\end{figure}

\end{subsubsection}

\begin{subsubsection}{modified $\phi^4$ lattice}

In the low temperature limit, $\langle x_i^2 \rangle \propto T^{2/3}$ is counter intuitive. To further understand the temperature dependence of $\langle x_i^2 \rangle$, we study a modified $\phi^4$ model. Its Hamiltonian reads
\begin{equation}      \label{eq.53}
  H = \sum_{i=1}^L \left[ \frac{p_i^2}{2m_i} + \frac{K}{2} (x_i-x_{i-1})^2 + \frac{\gamma}{4} x_i^4 + \frac{\theta}{2} x_i^2 \right].
\end{equation}
Here, a harmonic potential $(\theta/2) x_i^2$ is added. At $\theta = 0$, Eq.($\ref{eq.53}$) recovers the standard $\phi^4$ model.
Historically, Eq.($\ref{eq.53}$) with $\theta < 0$ (i.e., double potential well) has been used to study the structural phase transition\cite{SA53.1}.  The breather mobility\cite{DC53.3} and nonequilibrium statistical mechanical properties\cite{KA53.4} of this model were studied. In this work, we focus on the single potential well case ($\theta \ge 0$) and study the temperature dependence of $\langle x_i^2 \rangle $ and $\langle (x_i-x_{i+1})^2 \rangle$.
As to be shown below, a nonzero harmonic potential will change the low-temperature behavior of $\langle x_i^2 \rangle$.

In Figure $\ref{Fig6}$, we show the temperature dependence of $\langle x_i^2 \rangle$ for several values of $\theta$.
In the high temperature regime $T \gg T_{high}$, $\langle x_i^2 \rangle(T) \propto T^{1/2}$ and the coefficient is insensitive to $\theta$. This is because the $\gamma$ term dominates all the physical quantities in this limit. In the low temperature regime $T \ll T_{low}$, a finite $\theta$ leads to a new asymptotic behavior $\langle x_i^2 \rangle \propto T$. In the intermediate temperature $T_{low} \ll T \ll T_{high}$, $\langle x_i^2 \rangle \propto T^{2/3}$. The two crossover temperatures $T_{low}$ and $T_{high}$ are controlled by $\theta$ and $\gamma$, respectively. We find that $T_{low} \sim \theta^{3/2}$ and $T_{high} \sim \gamma^{-1}$. In thie limit $\theta=0$, $T^{2/3}$ behavior extends to $T=0$, recovering the result of Fig.$\ref{Fig5}(a)$. For $\theta \approx 1.0$, $T_{low} \approx T_{high}$ and the intermediate $T^{2/3}$ regime disappears.

The inset of Fig.$\ref{Fig6}$ shows $\langle (x_i - x_{i+1})^2 \rangle(T)$ for the same parameters. In contrast to $\langle x_{i}^{2} \rangle$, $\langle (x_i - x_{i+1})^2 \rangle(T)$ does not change qualitatively with $\theta$. At low temperature $T \ll T_{high}$, $\langle (x_i - x_{i+1})^2 \rangle \propto T$. At high temperature $T \gg T_{high}$, $\langle (x_i - x_{i+1})^2 \rangle \propto T^{1/2}$. Compared to $\langle x_i^2 \rangle$, $\langle (x_i - x_{i+1})^2 \rangle$ has only one crossover temperature $T_{high}$.

The rich physical behavior of Hamiltonian Eq.(\ref{eq.53}) is a result of competition between $U_h(x)$ (harmonic potential), $U_{ah}(x)$ (anharmonic potential), $V(x)$ (nearest-neighbor harmonic coupling), and temperature $T$ (kinetic energy).  $U_h(x)$ dominates the shape of the bottom of potential well and $U_{ah}(x)$ dominates the regime away from the bottom and brings the correlation. At low temperatures, the small-amplitude oscillation of particles is mainly constrained by $U_h(x)$ and $V(x)$, leading to $\langle x_i^2 \rangle \propto T$ and $\langle (x_i - x_{i+1})^2 \rangle \propto T$, respectively. In the high temperature limit, the large-amplitude and almost independent oscillations are dominated by $U_{ah}(x)$. So we have $\langle x_i^2 \rangle \sim \langle (x_i - x_{i+1})^2 \rangle \propto T^{1/2} $ according to Eq.(\ref{eq.49.4.3}). In the intermediate temperature region, the motion of the particles is jointly constrained by $U_h(x)$, $U_{ah}(x)$, and $V(x)$. The competition makes $\langle x_i^2 \rangle \propto T^{2/3}$, with the power index $2/3$ lying between $1$ ($U_h(x)$-dominant exponent) and $1/2$ ($U_{ah}(x)$-dominant exponent).

\end{subsubsection}

\end{subsection}

\end{section}


\begin{section}{Summary and Discussions}

In this work, we developed PTA in GF EOM formalism for classical systems. Using this method, we studied the one-dimensional $\phi^4$ lattice model under two successively larger variable bases. The result from one-dimensional B1 basis is found to be identical to that from the self-consistent phonon theory (equivalent to the quadratic variational method). The two-dimensional B2 basis gives quantitatively improved results. Qualitatively exact high- and low- temperature asymptotic behaviors for many static averages are obtained. The method presented in this work provides a new way to calculate the phonon spectrum of nonlinear lattice systems.

Our results in this work show that PTA is a systematic method to go beyond conventional variational method. There are many classical systems with interesting physical problems and are challenging for study. One of them is the FPU model that plays a central role in the study of low-dimensional anomalous heat transport problem and phonon transistor design\cite{NL44}. Another example is the molecular liquids, whose properties are very complicated and rich, especially close to the glass formation\cite{KN59,FA60}. The Coulomb fluid is still another example, where ions carrying positive and/or negative charges are dispersed in a liquid. The Coulomb interactions among ions could induce complicated phenomenon\cite{AN61}. In all these fields, PTA within GF EOM may be a useful tool.

The selection of basis is an important issue in PTA. In principle, we should incorporate in the basis those dynamical variables that are most relevant to eigen mode of the system. In practice, there are different ways to systematically enlarge basis. In this work, we start from the dynamical variable $Q_k$ (B1 basis). We then add $R_k$, which is a variable appearing in the expression $\{\{Q_k,H\},H\}$, to form B2 basis. We could collect the new variables appearing in $\{\{R_k,H\},H\}$ to further enlarge the basis. There are other strategies such as the Lanczos basis\cite{Lee1}. 
Different strategies of enlarging basis could have different converging speed. Finding an efficient basis selection strategy is an important issue to be studied in the future.

A related issue is the description of spectral function by PTA. 
Further study\cite{TM40.5} shows that while the static averages converge fast with increasing basis dimension, the spectral function depends more sensitively on the basis selection and converges slower. For classical systems, due to anharmonicity, the eigen mode frequencies depend on initial energy. Thermal averaging over initial states then broadens the excitation peaks at finite temperature, making them difficult to describe by a finite number of poles. As a result, for physical quantities that the damping of quasi-particles or the broadening of spectral function play a decisive role, such as transport coefficient, the present method may be inefficient. 
To overcome this difficulty, one could further calculate the memory function\cite{Mori,Zwanzig} contribution to self-energy based on PTA, along the line of Tserkovnikov\cite{Tserkovnikov1}. The spectral density approximation \cite{Kalashnikov1} may be advantageous in this regards, which assumes a continuous spectral function from the outset. 
Preliminary results with broadened spectral function have been obtained for $\phi^4$ lattice model and will be discussed elsewhere.

\end{section}

\vspace*{0.5cm}
\begin{section}{Acknowledgments }

This work is supported by the National Natural Science Foundation of China under Grant No. 11974420 (N. T.) and No. 12075316 (L. W.).
NHT is grateful to P. Zhang for helpful discussions.
Computational resources were provided by the Physical Laboratory of High Performance Computing at Renmin University of China.

\end{section}

\appendix{}

\section{Proof of fluctuation-dissipation theorem }

In this Appendix, we prove the fluctuation-dissipation theorem Eq.(\ref{eq.18.1}).
The proof was originally given in Ref.\cite{Herzel1}. Here we use a slightly different derivation.

The Fourier transformation of GF, Eq.(\ref{eq.15.1}), reads
\begin{eqnarray}      \label{eq.D1}
&& G^{r}(A|B)_{\omega} \nonumber \\
&=& \int_{-\infty}^{\infty}d(t-t^{\prime})  \theta(t-t^{\prime}) \langle \{ A(t),B(t^{\prime}) \} \rangle e^{i (\omega + i\eta) (t-t^{\prime}) }.  \nonumber\\
&&
\end{eqnarray}
Now we define the two-time correlation function $F\left[ A(t)|B(t^{\prime}) \right]$ as
\begin{equation}      \label{eq.D2}
  F\left[ A(t)|B(t^{\prime})\right] = \langle A(t) B(t^{\prime})\rangle.
\end{equation}
Its Fourier transformation $F(A|B)_{\omega}$ is given by
\begin{equation}      \label{eq.D3}
   F(A|B)_{\omega} = \int_{-\infty}^{\infty} d(t-t^{\prime}) F\left[ A(t)|B(t^{\prime}) \right] e^{i\omega(t-t^{\prime}) }.
\end{equation}
The EOM for $F\left[ A(t)|B(t^{\prime}) \right]$ reads
\begin{equation}      \label{eq.D4}
\frac{\partial}{\partial t} F\left[ A(t)|B(t^{\prime}) \right] = F\left[ \{A, H\}(t)|B(t^{\prime}) \right].
\end{equation}
The Fourier transformation gives
\begin{equation}      \label{eq.D5}
  \omega F(A|B)_{\omega} = i F(\{A,H\}| B)_{\omega}.
\end{equation}
It can be used to solve $F(A|B)_{\omega}$ except at $\omega=0$. For an example, for a constant variable $A=\alpha$, we have $\{A, H \}=0$ and the EOM reads
\begin{equation}      \label{eq.D6}
    \omega F(\alpha| B)_{\omega} = 0.
\end{equation}
It cannot be used to give the exact solution
\begin{equation}      \label{eq.D7}
F(\alpha| B)_{\omega} = 2\pi \alpha \langle B \rangle \delta(\omega).
\end{equation}

Using Eq.(\ref{eq.8.3}) obtained from the cyclic relation and the definition of $F\left[A(t)|B(t^{\prime})\right]$, Eq.(\ref{eq.D1}) becomes
\begin{eqnarray}      \label{eq.D8}
&& G^{r}(A|B)_{\omega} \nonumber \\
&& = \beta \int_{0}^{\infty} d(t-t^{\prime}) \left[\frac{\partial}{\partial t} F\left[ A(t)|B(t^{\prime}) \right] \right] e^{i (\omega + i\eta) (t-t^{\prime}) }.  \nonumber \\
&&
\end{eqnarray}
We then obtain
\begin{equation}      \label{eq.D9}
G^{r}(A|B)_{\omega} = \frac{\beta}{2 \pi} \int_{-\infty}^{\infty} d\omega^{\prime} \frac{\omega^{\prime} F(A|B)_{\omega^{\prime}}}{\omega + i \eta - \omega^{\prime}}.
\end{equation}
The corresponding Zubarev GF $G(A|B)_{\omega}$ reads
\begin{equation}      \label{eq.D10}
G(A|B)_{\omega} = \frac{\beta}{2\pi} \int_{-\infty}^{\infty} d\omega^{\prime} \frac{\omega^{\prime} F(A|B)_{\omega^{\prime}}}{\omega - \omega^{\prime}}.
\end{equation}
The spectral function $\Lambda_{A,B}(\omega)$ defined by Eq.(\ref{eq.19.1}) is obtained as
\begin{equation}      \label{eq.D11}
   \Lambda_{A,B}(\omega) = \frac{\beta}{2\pi} \omega F(A|B)_{\omega}.
\end{equation}
Its Fourier transformation is
\begin{equation}      \label{eq.D12}
  \Lambda_{A,B}(t-t^{\prime}) = \frac{i}{2\pi} \langle \{ A(t). B(t^{\prime} )\} \rangle
\end{equation}
At $t=t^{\prime}$, it gives the sum rule
\begin{equation}      \label{eq.D13}
  \int_{-\infty}^{\infty} d\omega \Lambda_{A,B}(\omega) = i \langle \{ A, B \} \rangle.
\end{equation}
Further, from Eq.(\ref{eq.D11}), one can obtain $F(A|B)_{\omega}$ as
\begin{equation}      \label{eq.D14}
   F(A|B)_{\omega} = \frac{2\pi}{\beta} \frac{\Lambda_{A,B}(\omega)}{\omega} + C_{A,B} \delta(\omega).
\end{equation}
$C_{A,B}$ is the contribution from zero-frequency components of $A(t)$ and$B(t^{\prime})$. It is given by
\begin{equation}     \label{eq.D15}
   C_{A,B} = 2 \pi \langle A_0 B_0 \rangle.  
\end{equation}   
Eqs.(\ref{eq.D14}) and (\ref{eq.D15}) are proved in Appendix B. 

Eqs.(\ref{eq.D14}) and (\ref{eq.D15}), together with the definition of $F\left[ A(t)| B(t^{\prime}) \right]$, give the fluctuation-dissipation theorem
\begin{equation}      \label{eq.D16}
 \langle A(t) B(t^{\prime}) \rangle = \frac{1}{\beta} \int_{-\infty}^{\infty} d\omega \frac{\Lambda_{A,B}(\omega)}{\omega} e^{-i\omega(t-t^{\prime})} + \langle A_0 B_0 \rangle.
\end{equation}
Eq.(\ref{eq.18.1}) in the main text is proved by taking $t=t^{\prime}$ in the above equation.

\section{Proof of the static contribution $C_{A,B} = 2\pi \langle A_0 B_0 \rangle$ }
In this Appendix, we discuss the properties of the zero-frequency component $X_0$ of an arbitrary variable $X$ in general, and prove Eqs.(\ref{eq.D14}) and (\ref{eq.D15}).

We first show that for a general variable $X(q,p)$, there is an unique splitting $X(q,p)=X_0(q,p) + \tilde{X}(q,p)$, in which $X_0(q,p)$ is the zero-frequency component and $\tilde{X}(q,p)$ is the nonzero-frequency component. Writing $X[x(t), p(t)]$ at time $t$ as $X(q,p;t)$ ($q=q(0)$ and $p=p(0)$), we can do a Fourier decomposition to it as
\begin{equation}     \label{B1}
   X(q,p;t) = X_0(q,p) + \sum_{n (\omega_n \neq 0)} X_n(q,p) e^{-i \omega_n t}.
\end{equation}
Here, $X_0(q,p)$ is the zero-frequency component of $X$ and the sum of other terms is the nonzero-frequency component. For nonperiodic system, replace the sum by integral and the following discussion still applies. Due to property (i) below, $X_0(q,p)$ is a conserved quantity and can be written as $X_0(q,p)=X_0[q(t),p(t)]$. Hence the nonzero-frequncy part $X[q(t),p(t)] - X_0[q(t), p(t)]$ is also a function of $q(t)$ and $p(t)$. This shows that for variable $X(q,p)$, we have a well-defined and unique splitting $X(q,p)=X_0(q,p) + \tilde{X}(q,p)$.

We have the following useful statements about $X_0$, $\tilde{X}$, and general conserved quantities.

(i) $\{X_0, H\} = 0$.

This is easily obtained from $\partial X_0(q,p)/ \partial t = 0 = \{ X_0, H\}(q,p)$. It shows that the static component of any variable $X$ is a conserved quantity. 

(ii) $\langle \{X_0, O\} \rangle =0$ for any variable $O$.

This can be obtained from $\langle \{X_0, O \} \rangle = \beta \langle \{X_0, H\} O \rangle = 0$. Here property (i) is used. It follows from (ii) that $X_0$ is orthogonal to any dynamical variable $O$ under the inner product Eq.(\ref{eq.36.5}). That is, $(X_0|O) =  0$. To show this, we substitute $O$ in (ii) by $\{O^{\ast}, H \}$, take a complex conjugate, and assume that $H^{\ast} = H$. In particular, $(X_0|X_0) =0$, meaning $X_0$ has zero length under the inner product Eq.(\ref{eq.36.5}).

(iii) The space of conserved variables is the space of zero length variables. 

On one hand, a conserved variable $A$  fulfils $\{A, H\} =0$ by definition. We have $(A|A) = \langle \{A^{\ast}, \{A, H\} \} \rangle = 0$, i.e., $A$ has length zero. On the other hand, if a variable $A$ has zero length, i.e., $(A|A)=0$, we have $ 0 = \langle \{A^{\ast}, \{A, H\} \} \rangle = \beta \langle \{A^{\ast}, H\} \{A, H\} \rangle$. It implies $\{ A, H \} =0$. Mathematically, this space is called the null space of the Liouville operator $\mathcal{L} = \{ ..., H\}$.

(iv) $\langle A_0 \tilde{B} \rangle = \langle \tilde{A} B_0 \rangle =0$ for any two variables $A$ and $B$.

From Eq.(\ref{B1}), we can write $\langle A_0 \tilde{B} \rangle = \sum_{n(\omega_n \neq 0)} \langle A_0 B_n \rangle$, employing the time translational invariance of the equilibrium state.  
We also have $\partial \tilde{X}(q,p;t) / \partial t = \sum_{n (\omega_n \neq 0)} \{ X_n(q,p), H \} e^{-i\omega_n t} = \sum_{n (\omega_n \neq 0)} (-i\omega_n)  X_n(q,p) e^{-i\omega_n t}$. It gives $\{ X_n(q,p), H \} = -i\omega_n X_n(q,p)$.
Applying it to $B_n$, we have $\langle A_0 \tilde{B} \rangle = \sum_{n(\omega_n \neq 0)} (-1/ i\omega_n) \langle A_0 \{B_n, H\} \rangle= \sum_{n(\omega_n \neq 0)} (1/i\omega_n) \langle \{A_0, H\} B_n \rangle = 0$. Similarly, $\langle \tilde{A} B_0 \rangle =0$. In particular, let $B=c \neq 0$ being a constant, we obtain $\langle \tilde{A} \rangle = 0$ for arbitrary variable $A$.

Having obtained these properties, we consider the relaxation function $F[A(t)|B(t^{\prime})]$ defined in Eq.(\ref{eq.D2}). Now we have $F[A(t)|B(t^{\prime})] = \langle \left[ A_0 + \tilde A(t) \right] \left[ B_0 + \tilde B(t^{\prime}) \right]  \rangle$. Using (iv) and doing Fourier transformation, we obtain
\begin{equation}   \label{B2}
   F(A|B)_{\omega} = F(\tilde{A}|\tilde{B})_{\omega} + 2\pi \langle A_0 B_0 \rangle \delta(\omega).
\end{equation}
Eq.(\ref{eq.D11}) gives
\begin{equation}   \label{B3}
    F(\tilde{A}|\tilde{B})_{\omega} = \frac{2\pi}{\beta \omega} \Lambda_{\tilde{A}, \tilde{B}}(\omega).
\end{equation}
From the definitions of retarded Green's function $G^{r}\left[A(t)|B(t^{\prime}) \right]$ in Eq.(\ref{eq.11.1}) and the spectral function $\Lambda_{A,B}(\omega)$ in Eq.(\ref{eq.19.1}), using (ii), we can easily confirm that both depend only on the nonzero-frequency components of $A$ and $B$. That is, $G(A|B)_{\omega} = G(\tilde{A}| \tilde{B})_{\omega}$ and $\Lambda_{A,B}(\omega) = \Lambda_{\tilde{A}, \tilde{B}}(\omega)$. Putting the latter equation into Eqs.(\ref{B2}) and (\ref{B3}), we obtain
\begin{equation}   \label{B4}
    F(A|B)_{\omega} = \frac{2\pi}{\beta \omega} \Lambda_{A, B}(\omega)+ 2\pi \langle A_0 B_0 \rangle \delta(\omega).
\end{equation}
This completes the proof of Eqs.(\ref{eq.D14}) and (\ref{eq.D15}).

\end{document}